\documentstyle[twocolumn,epsf]{jpsj}

\catcode`\@=11

\def\simle{\mathrel{\mathpalette\@versim<}}   
\def\simge{\mathrel{\mathpalette\@versim>}}   
\def\@versim#1#2{\lower2.5pt\vbox{\baselineskip0pt \lineskip-.5pt
   \ialign{$\m@th#1\hfil##\hfil$\crcr#2\crcr\sim\crcr}}}

\catcode`\@=12

\def\runtitle{
}
\def\runauthor{Yukitoshi {\sc Motome} and Nobuo {\sc Furukawa}}

\def\sf{\rm}

\title{
Non-equilibrium Relaxation Study of Ferromagnetic Transition\\
in Double-Exchange Systems
}

\author{
Yukitoshi {\sc Motome} and Nobuo {\sc Furukawa}$^{1}$ 
}
\inst{
Institute of Materials Science, University of Tsukuba,
Tsukuba, Ibaraki 305-0006 \\
$^{1}$Department of Physics, Aoyama Gakuin University, 
Setagaya, Tokyo 157-8572
}
\recdate{
}

\abst{
Ferromagnetic transition in double-exchange systems is studied by non-
equilibrium relaxation technique combined with Monte Carlo calculations. 
Critical temperature and critical exponents are estimated from 
relaxation of the magnetic moment. The results are consistent with the 
previous Monte Carlo results in thermal equilibrium. The exponents 
estimated by these independent techniques suggest that the universality 
class of this transition is the same as that of short-range interaction 
models but is different from the mean-field one.
}

\kword{
double-exchange system, ferromagnetic transition, non-equilibrium relaxation,
Monte Carlo, polynomial expansion, colossal-magnetoresistance manganites
}

\begin{document}
\sloppy
\maketitle


Since the proposal by Zener,
\cite{Zener1951}
the double-exchange (DE) mechanism has been studied
to explain ferromagnetic transition in doped manganites
which show the colossal magnetoresistance.
The ferromagnetism itself is well explained by the DE mechanism;
at low temperatures, itinerant electrons which interact with localized spins
through the strong Hund's-rule coupling
favor parallel configuration of localized spin
in order to gain the kinetic energy.
The colossal magnetoresistance is understood by this mechanism also,
at least, qualitatively.

However, in spite of the long history of theoretical studies,
finite-temperature properties of the DE systems
have not been fully understood thus far.
At finite temperatures, especially near the ferromagnetic transition,
fluctuations are quite large
through the strong interplay between itinerant electrons and localized spins.
This makes it difficult to describe the thermodynamics quantitatively.
Approximational methods, for instance, the mean-field approximation,
have been known to be insufficient because they neglect the fluctuation effects.
\cite{Furukawa1999}

Recently, the thermodynamics of the DE systems has been studied intensively
by using the Monte Carlo (MC) method which fully includes large fluctuations.
\cite{Motome1999,Motome2000,Motome2001}
A new algorithm is developed by using the moment-expansion technique
for the density of states to calculate the MC weight efficiently and
the computational speed is accelerated considerably
by using parallel computers.
This enables us to calculate large-size clusters and
take account of the finite-size errors systematically by the finite-size scaling.
The critical temperature $T_{\rm c}$ and the critical exponents are estimated
by applying the finite-size scaling analysis on the MC data.

The previous MC studies are for thermal equilibrium.
Some initial MC steps (typically 1,000-10,000 steps) are discarded
for thermal equilibration, and
measurement of physical properties is performed
by taking the thermal averages for the grand canonical ensemble
using MC samples after the thermalization.

There is another powerful technique to investigate the critical properties
by using the MC method, i.e., the non-equilibrium relaxation (NER) technique.
\cite{Kikuchi1986,Stauffer1992,Kohring1992,Ito1993}
In this technique, dependence on MC steps of the order parameter is studied
in the non-equilibrium relaxation
from an initial state which is chosen to be a symmetry-broken one.
The order parameter decays exponentially
above the critical temperature $T > T_{\rm c}$.
On the other hand, it approaches a constant for $T < T_{\rm c}$.
As a critical relaxation, a power decay is observed at $T = T_{\rm c}$
whose exponent relates to the critical exponents.

One of the advantages of the NER technique is that
one can study larger-size clusters than by the equilibrium technique
since the relaxation is measured in the primary part of the MC samplings
which are discarded as a thermalization process in the equilibrium method.
The finite-size effect can be exponentially small and neglected in the relaxation
if one calculates a sufficiently large-size cluster.
The NER technique gives information on the critical temperature and
the critical exponents, independently of the equilibrium MC method.

In this paper, we investigate the ferromagnetic transition in DE systems
by the NER technique combined with the MC calculations.
To our knowledge, this is the first application of the technique
to critical phenomena in itinerant electron systems.
The results are compared with those by the previous equilibrium technique.

We study the DE model where itinerant electrons couple to
localized spins with Ising symmetry on square lattices.
This is a minimum model which shows the ferromagnetic transition
at finite temperatures by the DE mechanism.
We consider the limit of strong Hund's-rule coupling for simplicity.
The Hamiltonian is written by
\cite{Motome2001}
\begin{equation}
\label{eq:H}
{\cal H} = -  \sum_{<ij>} \frac{t}{2}(1 + S_{i}S_{j}) \ 
c^{\dagger}_{i} c_{j},
\end{equation}
where $c_{i}$ ($c^{\dagger}_{i}$) annihilates (creates)
an electron at site $i$, and
$S$ describes the localized Ising spin which takes $S=\pm 1$.
The summation is taken for nearest-neighbor sites.

We apply the moment-expansion MC method.
\cite{Motome1999}
We take the same computational conditions
as in ref.~[\citen{Motome2001}].
We study a square lattice of $24 \times 24$ sites
which seems to be large enough to observe the critical relaxation.
We will comment on the finite-size effect later.
The relaxation is measured for typically 100-1000 different
Markov sequences to estimate statistical errors.
The energy unit is the half-bandwidth of noninteracting electrons; $W = 4t$.

\begin{figure}[htb]
\epsfxsize=8.cm
\centerline{\epsfbox{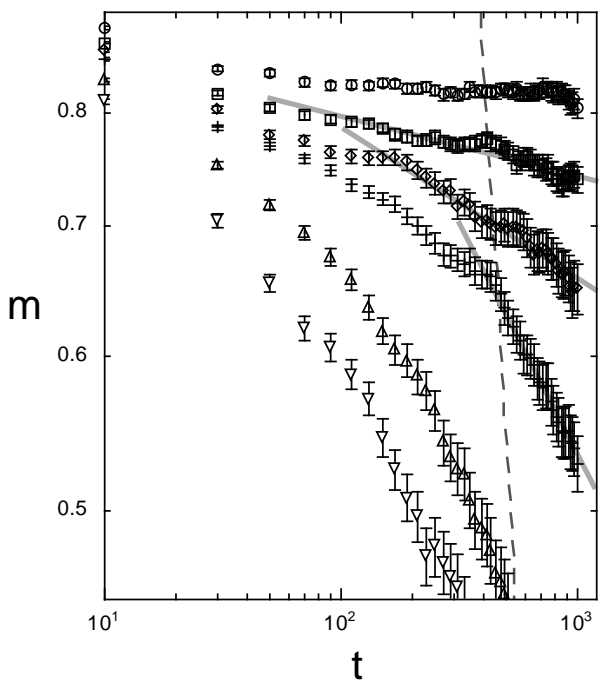}}
\caption{
Relaxation of the magnetic moment $m$ in Monte Carlo steps $t$
from the ferromagnetic state with perfect polarization.
The data are for $T/W = 0.056, 0.057, 0.0575, 0.058, 0.059$ and $0.06$
from top to bottom, respectively.
The gray lines are the least-squares fit
by power functions for the asymptotic behavior
of the data at $T/W = 0.057, 0.0575$ and $0.058$.
The dotted line is the asymptotic power decay 
in the mean-field universality class, $m \sim t^{-2}$, for comparison.
}
\label{fig:relaxation}
\end{figure}

We calculate the relaxation of the magnetic moment per site
from the perfectly polarized configuration of localized spins.
Figure~\ref{fig:relaxation} shows the results.
For $T/W \simge 0.059$, the moment decays exponentially.
This indicates that the system is in the paramagnetic state, i.e., $T > T_{\rm c}$.
On the contrary, the data at $T/W = 0.056$ seem to approach a constant,
which suggests the broken symmetry below $T_{\rm c}$.
Power-decay behavior is observed between these temperatures.
We estimate $T_{\rm c}/W = 0.0575 \pm 0.001$.
This result is consistent with the estimate
by the equilibrium technique,
$T_{\rm c}/W = 0.058 \pm 0.001$.
\cite{Motome2001}

The asymptotic exponent of the power decay gives
information on the critical exponents.
The magnetic moment $m$ depends on the MC step $t$
as $m \sim t^{-\lambda}$ in the limit of $t \rightarrow \infty$,
where $\lambda = \beta / \nu z$.
Here, $\beta$ and $\nu$ are the exponents 
for the order parameter and the correlation length, respectively, 
and $z$ is the dynamical exponent.
We fit the asymptotic tails of the data
at $T/W = 0.057, 0.0575$ and $0.058$ by power functions
as shown by the gray lines in Fig.~\ref{fig:relaxation} and 
obtain the estimate $\lambda = 0.14 \pm 0.1$.

The estimate of $\lambda$ is examined to identify
the universality class of this transition.
The mean-field universality class gives the exponents as
$\beta = \nu = z = 1/2$.
The dynamical exponent $z$ is obtained from the consideration that
the MC update from $n$ to $n+1$-th step at $T = T_{\rm c}$ is described by
$m_{n+1} = \tanh(m_{n})$ in mean-field models.
Thus, the exponent $\lambda$ is $2$ in the mean-field universality class.
On the other hand, in the universality class of the Ising model
with short-range interactions in two dimensions,
the critical exponents are estimated
as $\beta = 1/8$, $\nu = 1$
\cite{Kaufmann1944,Stephenson1964}
and $z = 2.16$.
\cite{Ito1993}
In this case, the exponent $\lambda$ becomes $0.058$.
Our estimate $\lambda = 0.14 \pm 0.1$ is consistent with
the universality class of the short-range Ising model, and
is distinct from the mean-field one.

The present results are compared with the previous ones
obtained by the equilibrium MC calculations.
\cite{Motome2001}
The previous estimates of the critical exponents were
$\beta = 0.09 \pm 0.08$ and $\nu = 0.9 \pm 0.3$, which are
consistent with those for the short-range Ising model
but different from the mean-field ones.
Therefore, the present result by the NER technique
is consistent with those by the equilibrium technique.
It is suggested that the universality class of the ferromagnetic transition
in the DE system is the same as that of short-range interaction models
with the same spin symmetry, but different from the mean-field one,
from these independent techniques.

We comment on the finite-size effect in the present NER study briefly.
One of the finite-size effects is an exponential decay
due to an energy gap inherent in finite-size systems.
This should decrease the temperature where the relaxation
shows the power-decay behavior.
Thus, the present estimate of $T_{\rm c}$ is considered to give a lower limit.
The agreement between the present estimate and the previous one
by the equilibrium technique indicates that
the finite-size effect is negligibly small
in the present precision of the data.

To summarize, we have investigated the ferromagnetic transition
in the double-exchange systems
by the non-equilibrium-relaxation Monte-Carlo method.
To our knowledge, this is the first example of the application of this method
to the phase transition in itinerant electron systems.
From the relaxation of the magnetic moment,
we have estimated the critical temperature and the critical exponents.
The results are consistent with the previous ones
which are obtained by the finite-size scaling analysis on
the Monte-Carlo data in thermal equilibrium.
The exponents estimated by these independent techniques
indicate consistently that the universality class of this transition
appears to be the same as that of short-range interaction models
but is different from the mean-field one.


The authors thank H. Nakata for helpful support
in developing parallel-processing systems.
The computations have been performed mainly 
using the facilities in the AOYAMA+ project
(http://www.phys.aoyama.ac.jp/\\ \~{}aoyama+)
and in the Supercomputer Center, Institute for Solid State Physics,
University of Tokyo.
This work is supported by  ``a Grant-in-Aid from
the Ministry of Education, Culture, Sports, Science and Technology''.



\begin{thebibliography}{99}

\bibitem{Zener1951}
C. Zener:
Phys. Rev. {\bf 82} (1951) 403.

\bibitem{Furukawa1999}
N. Furukawa:
in {\em Physics of Manganites}, 
eds. T. Kaplan and S. Mahanti (Plenum Publishing, New York, 1999),
and references therein.

\bibitem{Motome1999}
Y. Motome and N. Furukawa:
J. Phys. Soc. Jpn. {\bf 68} (1999) 3853.

\bibitem{Motome2000}
Y. Motome and N. Furukawa:
J. Phys. Soc. Jpn. {\bf 69} (2000) 3785.

\bibitem{Motome2001}
Y. Motome and N. Furukawa:
J. Phys. Soc. Jpn. {\bf 70} (2001) No.6 in press.

\bibitem{Kikuchi1986}
M. Kikuchi and Y. Okabe:
J. Phys. Soc. Jpn. {\bf 55} (1986) 1359.

\bibitem{Stauffer1992}
D. Stauffer:
Physica A {\bf 186} (1992) 197.

\bibitem{Kohring1992}
G. A. Kohring and D. Stauffer:
Int. J. Mod. Phys. C {\bf 3} (1992) 1165.

\bibitem{Ito1993}
N. Ito:
Physica A {\bf 192} (1993) 604.

\bibitem{Kaufmann1944}
B. Kaufmann and L. Onsager:
Phys. Rev. {\bf 76} (1944) 1244.

\bibitem{Stephenson1964}
J. Stephenson:
J. Math. Phys. {\bf 5} (1964) 1009.

\end{thebibliography}
\end{document}